\documentstyle[aps,preprint]{revtex}
\begin{document}
\draft
\title{Confinement and death of oscillations in coupled chaotic bistable oscillators}
\author{Hwa-Kyun Park\footnote{e-mail: childend@front.kaist.ac.kr \\
 TEL:+49-351-871-1225 \hspace{.5cm} FAX: +49-351-871-1999} .}  
\address{Department of Physics,Korea Advanced Institute of Science and
  Technology,Taejon 305-701,Korea} 
\address{ Max-Planck Institut f\"ur Physik komplexer Systeme,
Dresden 01187, Germany }
\date{\today}

\maketitle

\begin{abstract}
In coupled chaotic bistable systems such as Lorenz and Chua  
oscillators, two-phase domains corresponding to the two lobes
of the strange attractor are formed. 
The dynamics of each domain is confined to one lobe
and typically exhibits one of the two types of behavior: oscillation
death or nearly periodic oscillations.
We elucidate the role of intrinsic broad time scales on 
the confinement and the oscillation death.
\end{abstract}

\pacs{PACS numbers: 05.45.Xt, 47.54.+r}

Many systems in physics, chemistry, and biology can be
considered as assemblies of coupled oscillators
\cite{winfree,kuramoto}.
These systems can be categorized by the local dynamics
of oscillators and the network structures between them.
In reaction diffusion systems, oscillators are coupled
diffusively, and the local dynamics can be bistable, excitable
or oscillatory.
They are classified according to their local dynamics and
the characteristics of each class have been well understood
\cite {mik}.

In recent years, cases where the local dynamics is
chaotic have aroused a great deal of interest.
Studies were performed using various models such as
coupled map lattices \cite{cml,cml2,ntcb}, and coupled oscillators 
\cite{ntcb2,size,line,clorenz1,clorenz2,death2}.
Synchronization \cite {synch,synch2,synch3} is 
the most important feature of two coupled chaotic oscillators.
However $N$ coupled chaotic oscillators with large $N$
exhibit not only synchronization but also more complex behaviors
such as nontrivial collective behavior \cite{ntcb,ntcb2}
and size instability \cite{size}.
In particular, it has been revealed that from topological considerations
line defects are a general property of coupled oscillators with
period doubling bifurcations \cite{line}, and such line defects 
have been observed in a recent experiment \cite{line2}.

However there are many other chaotic oscillators which do not
follow period doubling bifurcations \cite{chaos}
and more studies on coupled systems of such oscillators are needed. 
Especially we want to know whether universal properties
exist and can be categorized by the characteristics
of the local dynamics. 
In this paper, we investigate the dynamics of the $N$ coupled 
chaotic bistable systems.
We call systems like the Lorenz, Chua, and 
Duffing oscillators ``chaotic bistable systems'' \cite{synch3}. 
In these systems, two bistable basins exist below some threshold, 
and chaotic orbits wandering between two regions emerge above 
this threshold.

To begin with, we consider the following $N$ 
diffusively coupled Lorenz oscillators with periodic boundary
conditions,

\begin{eqnarray}\label{eq-lorenz}
{dx_{i}} \over {dt} &=& \sigma (y_{i}-z_{i})+\epsilon (x_{i+1}+x_{i-1}-2 x_{i}) \\ 
{dy_{i}} \over {dt} &=& r x_{i}-y_{i}-x_{i}z_{i}+\epsilon (y_{i+1}+y_{i-1}-2 y_{i}) \nonumber \\
{dz_{i}} \over {dt} &=& x_{i}y_{i}-bz_{i}+\epsilon (z_{i+1}+z_{i-1}-2z_{i}). \nonumber
\end{eqnarray}
We will concentrate on the canonical parameter values $\sigma=10,b=8/3$ and
$r=28$ \cite{lorenz}.
The uncoupled Lorenz oscillator shows a chaotic
trajectory consisting of two lobes.
The trajectory circulates around a right saddle-focus a few times,
moves to the left, and circulates around a left saddle-focus, and 
so on. These alternations are repeated indefinitely and erratically.

One thing to note for later reference is that the Lorenz oscillator has
broad time scales \cite{multiple}. Characteristics of a chaotic
attractor can be well resolved by unstable periodic orbits(UPO)
\cite{upo} and UPOs of Lorenz oscillators show various periods 
\cite{multiple}.
Another method to examine the broad time scales is to calculate the 
instantaneous frequency $\omega(t)$. For Lorenz systems, the phase 
variable $\phi$ can be defined by
$\tan^{-1} (z-z_0)/(A-A_0)$ where $z_0=r-1$, $A=\sqrt{x^2+y^2}$,
$A_0=\sqrt{2b(r-1)}$ \cite{phase} and then
$\omega(t)=\dot {\phi}(t)$ is given by 
\begin{equation}\label{eq-w}
\omega(t)=\frac{A(A-A_0)\dot{z} - (z-z_0)(x \dot{x}+ y \dot{y})}
{A((z-z_0)^2+(A-A_0)^2)}.
\end{equation}
Fig. \ref{local}(a) shows the time evolution of $\omega(t)$ obtained 
from a typical Lorenz attractor.
Compared with the R\"{o}ssler system (Fig. \ref{local}(b)),
it clearly shows a broad distribution of $\omega(t)$.
Note that the scale of ordinates is different for Lorenz 
and R\"{o}ossler system in Figs. \ref{local}(a),(b) considering
their difference of mean frequencies.

For coupled Lorenz systems, steady states, synchronized 
oscillations \cite{clorenz1}, and front motions \cite{clorenz2}
have been studied.  In particular, a linear stability analysis 
shows the existence of a large number of steady states
\cite{clorenz1}. 
However it has not been clearly explained what characteristics of the
system cause such dynamics. The purpose of the present work is
to suggest the underlying physical mechanism and confirm its
universality.

While the synchronization of a small number of coupled Lorenz
oscillators is achieved with weak coupling,
this is not the case for a large number of coupled oscillators
\cite{size}.
In particular, a linear stability analysis shows that
the Lyapunov exponent of the {\it k}th Fourier component 
$\lambda_k$ is given by 
$\lambda_k=\lambda_0-4\epsilon \sin^{2}(\pi k /N)$ \cite{size}, 
where $\lambda_0$ corresponds to the Lyapunov exponent of
the uncoupled single oscillator.
Therefore for a fixed coupling constant $\epsilon$, as the system 
size $N$ increases beyond some critical value, synchronous states 
become unstable. 
 
Shown in Figs. \ref{coupled}(a) and (c) are numerical results of
Eq. (\ref{eq-lorenz}). 
Two-phase domains corresponding to the left lobe and the right 
form. In other words, the trajectory of each oscillator is confined
either to the left lobe or the right in contrast to the uncoupled
single oscillator.
The domains show two types of dynamics: 
frozen or pulsating, as in Figs. \ref{coupled}(a),(c). 
In a frozen state, oscillations are suppressed to stationary 
states. However pulsating domains show recurrent 
oscillations \cite{recur}.  Note that frozen and pulsating 
domains are coexisting. 

Now we will focus on the mechanism of such dynamics.
As explained above, the Lorenz system has broad time
scales.
In Eq. (\ref{eq-lorenz}), $-2\epsilon x_i$,$-2\epsilon y_i$ 
and $-2\epsilon z_i$ provides
some additional damping, however, the forcing from the neighboring
oscillators has minor effects since it is nonresonant due to the
broad time scales. Therefore damping is dominant and  
oscillations are suppressed.

Although we deal with the identical oscillators,  
these two types of domains can be well understood using the
terminology of coupled nonidentical oscillators.
It has been reported that coupled nonidentical
oscillators with various frequencies \cite{death,death2}
exhibit very rich dynamics including steady states such as 
oscillation death, as well as unsteady states such as 
periodic, quasiperiodic or chaotic evolution.
Although we consider the identical Lorenz oscillators,
they have broad time scales, so effectively they act as  
oscillators with various natural frequencies \cite{multiple}.
We can then view the frozen states of coupled Lorenz systems as
the oscillation death, and the pulsating states as the 
unsteady states of coupled oscillators of various frequencies.

In coupled periodic oscillators\cite{death} or chaotic oscillators
such as R\"{o}ssler systems\cite{death2}, for the forcing from 
the neighbor to be nonresonant so that the oscillation death occurs, 
coupling with nonidentical oscillator is required.
However, for the Lorenz systems, oscillation death can be
obtained from the coupling of the identical oscillators
since they have broad time scales.

For a more quantitative study, the local frequency gradient, 
$\delta w=(|\omega_{i}-\omega_{i-1}|+|\omega_{i}-\omega_{i+1}|)/2$,
is calculated from the same data of Fig. \ref{coupled}(a), 
and plotted in Fig. \ref{coupled}(b).   
We want to emphasize that $\omega$ is defined in the absence 
of coupling to reflect the intrinsic property of the local dynamics
and given by Eq. (\ref{eq-w}) which is derived from the equations
of the uncoupled single Lorenz oscillator.
Compare Fig. \ref{coupled}(a) with Fig. \ref{coupled}(b), and see
Fig. \ref{coupled}(d) and (e), 
where $\omega_{i}$ are obtained for the regions denoted by 
``frozen'' and ``pulsating'' in Fig. \ref{coupled}(a), respectively.
From these results, one can see that the oscillation death
corresponds with large frequency mismatches between neighboring 
sites, and oscillating state with the small mismatches.
Note that small clusters with close frequencies in Fig. 2(d)
correspond to the boundary of domains where simple comparisons of 
frequencies between neighboring sites are not sufficient for the 
analysis of dynamics.  

The relation between the dynamics of domains and the frequency
mismatches is more clearly shown in Fig. \ref{hist}, where
probability distribution function  of frequency mismatches,
$\delta w$, of space-time points for frozen and pulsating 
domains are denoted as solid and dashed lines respectively
To get the histograms in Fig. \ref{hist}, 100 independent
numerical simulations with different initial conditions 
are considered.  Initial transients ($t<300$) are excluded, 
and data are obtained with a time step $\Delta t=0.025$  
for each space point during next 50 sec. Also, boundary
points of domains are excluded.  Fig. \ref{hist} shows 
that the probability distribution function for the pulsating 
domains is peaked at lower values of frequency mismatch
compared with that of the frozen domains.

The results are consistent with those of previous studies on the coupled
nonidentical chaotic oscillators\cite{death2}.  In that case, it was
reported that the large frequency mismatches induce the oscillation 
death and the small ones bring the synchronized oscillation.

Rigorously speaking, the exactly stationary states cannot
coexist with the oscillating states in the same lattice. 
But the effect from neighboring pulsating domains are
so well blocked that the frozen domains can be essentially 
stationary except at the boundary between the frozen and 
pulsating domains.  For the frozen domains, we could not
find any noticeable oscillations numerically.  

Also note that bistability of the local dynamics is important 
for oscillation death. Numerical results show that larger 
domains tend to be pulsating rather than frozen. It means that 
there is an upper limit for the width of the frozen domain.
Without bistability, there could be no domain structure, and
so oscillation death could not occur.

Next, we investigate the effects of forcing on the coupled Lorenz 
systems, which confirm the above explained role of broad time scales 
again.
For the preservation of the mirror symmetry of the system, 
a force $A\cos(\Omega t)$ is applied to the z variable.
For uncoupled systems, synchronization with the external periodic 
forcing is imperfect due to the broad distribution of 
the intrinsic time scale \cite{multiple}.
Here, synchronization is defined by the absent or negligible difference
between the external forcing frequency  $\Omega$ and the average phase
velocity $\overline{\omega}$ of the system obtained from a sufficiently
long time series. For a fixed $A$, $\overline{\omega} - \Omega$ is
calculated as a function of $\Omega$; typical results are presented in
Fig. \ref{forced}(a). 
There exists a horizontal plateau with some fluctuations near zero.
This imperfect synchronization region as a function of the external 
forcing frequency $\Omega$ and amplitude $A$ is depicted by solid 
lines in Fig. \ref{forced}(b).

For coupled systems with time-varying forcing, clearly, 
oscillation death is impossible. So we concentrate on the effects
of forcing on the confinement.
The external force entrains the system, making a characteristic
time scale. Hence the chaos suppression effect, which was enhanced by
the broad time scales, is weakened. Thus the resonant external force
breaks the confinement. 
As a function of $\omega$ and $A$, two regions (I and II) are obtained 
in Fig. \ref{forced}(b). In Region I (shaded region), confinement is 
broken and the dynamics of the system becomes turbulent 
(Figs. \ref{forced}(d),(f));
In Region II (unshaded region), one can see two-phase domains with 
confined dynamics (Fig. \ref{forced}(e)).  Region I coincides with 
the region of synchronization with external forcing of the 
uncoupled oscillator.

However there are some deviations: Region I is shifted
to higher frequencies compared with the synchronization of the local
dynamics. In fact, the situation is very similar to the results 
from the complex Ginzburg Landau equation with external 
forcing where coupling changes the frequency of the system, so that
Arnold tongue structure changes compared with the uncoupled cases
\cite{cgl}.
The UPOs of the Lorenz system typically consist of 
fast rotations in one lobe and slow transition to another lobe
\cite{multiple}.
With coupling, the long time transition between two lobes is
prohibited and effectively the frequency of system is increased,
which may explain the above described deviation. 
When the confinement is broken by the external force, the prevented
transitions are possible and the dynamics of the system becomes slow. 
Then the system can be nonresonant to the external force and 
confined again.
Thus, at the boundary between Region I and Region II, one can 
see mixed patches of domain structure where domains with the
confinement and without the confinement are coexisting 
as in Fig. \ref{forced}.(f).

Thus far, we have shown that the broad time scales induce oscillation
death and confinement in the coupled Lorenz oscillators.
To confirm the universality of these results, we now turn to the
dynamics of coupled Chua \cite{chua} and periodically forced
Duffing oscillators. 

For the coupled Chua oscillators, we consider the
dynamics of the following equations:
\begin{eqnarray}
{dx_{i}} \over {dt} &=&  c_1 (y_{i}-x_{i}-g(x_{i})) + 
\epsilon (x_{i+1}+x_{i-1}-2x_{i}) \\
{dy_{i}} \over {dt} &=& 
c_2 (x_{i}-y_{i}+z_{i}) + \epsilon (y_{i+1}+y_{i-1}-2y_{i}) \nonumber \\
{dz_{i}} \over {dt} &=&  -c_3 y_{i} + \epsilon (z_{i+1}+z_{i-1}-2z_{i}),
\nonumber
\end{eqnarray}
where $g(x)=m_1 x + \frac{m_0-m_1}{2}(|x+1|-|x-1|)$ 
($c_1=15.6$,$c_2=1$,$m_0=-8/7$, $m_1=-5/7$,$c_3=24$,$\epsilon =5.0$).

The single Chua oscillator without coupling yields a chaotic attractor
with two wing structure as in Fig. \ref{chua}(a).
To check the broadness of the time scales, we should define the
phase and calculate the instantaneous frequencies. However,
in this case, a unique center of rotation is not well defined, so
a simple formula for the instantaneous frequency as a function of
arbitrary $x,y,z$ can not be obtained. We define the phase
for the Chua oscillator by the empirical-mode-decomposition 
method\cite{phase2}, and then the instantaneous phase velocity 
shows broad time scales as in Fig. \ref{chua}(b). 
The dynamics of the coupled Chua oscillators is quite similar to
that of the coupled Lorenz oscillators. For weak coupling,
we have imperfect confinement and no oscillation
death as in Fig. \ref{chua2}(a). However, with increasing
forcing strengths, confined domains show two types of motions
: oscillation death or nearly periodic oscillations as in the
case of Lorenz oscillators.
As seen in Figs. \ref{chua2}(b) and (c), frozen domains and oscillating
domains coexist. 

The equations for the lattice of forced Duffing oscillators are
given by
\begin{eqnarray}
{dx_{i}} \over {dt} &=& y_{i}+\epsilon (x_{i+1}+x_{i-1} - 2 x_{i}) \\
{dy_{i}} \over {dt} &=& a x_{i}^3 + b x_{i} + c y_{i} + f \sin(\omega t) +
\epsilon (y_{i+1}+y_{i-1}- 2 y_{i}) \nonumber 
\end{eqnarray}.
The orbit of a single forced Duffing oscillator shows a chaotic bistable 
nature as in Figs. \ref{duffing}(a),(c).
But the situation is different with the Lorenz oscillator or the
Chua oscillator.
For the lattice of forced Duffing oscillators, oscillation death
cannot be obtained due to the external time-varying force.
Also, confinement is expected to be easily broken since the
time scales of dynamics are dominated by the external forcing 
frequency. 
Since we are interested only in the confinement phenomena,
we use the binary representation of space time evolutions
of dynamics: white(black) denotes positive(negative) x
in Figs. \ref{duffing} (b),(d),(e),(f).
For weak forcing amplitudes, confinement is obtained 
with weak couplings (See Fig. \ref{duffing}(b)).
But boundaries between the domains are oscillating.
With increasing coupling constant $\epsilon$, width of domains increase, 
however the boundary instability does not disappear.
For strong forcing amplitudes, it is difficult to
obtain the confinement. For lower coupling strengths,
spatiotemporal chaotic pattern occurs as in Fig. \ref{duffing}(d). 
With increasing diffusive coupling constant $\epsilon$ (about more 
than $\epsilon=50$), we get confined domains, however, confinement is 
still imperfect at the boundaries of domains as in Fig. \ref{duffing}(e).
Further increasing coupling strength, the domain structure disappears
and the confinement is broken as in Fig. \ref{duffing}(f).
For sufficiently large coupling strengths, synchronization of whole
system occurs.

In summary, we have investigated the mechanism of confinement and 
oscillation death in coupled chaotic bistable systems.
With the broad time scales of the local dynamics, diffusive
coupling suppresses the oscillation and causes confinement
and oscillation death.
The result implies that in coupled chaotic oscillators,
the general properties of dynamics can be classified by 
the characteristics of the individual oscillators.
The author wishes to thank H.T.Moon, S.-O.Jeong and T.-W.Ko for many
stimulating discussions and M. B\"aer for useful comments.
This work was supported in part by the interdisciplinary research
(Grant No.1999-2-112-002-5) program of the KOSEF and
in part by the BK21 program of the Ministry of Education in Korea.

\pagebreak
\newpage


\begin{figure}
\caption{
Instantaneous frequencies  $\omega(t)$ obtained from (a) Lorenz system
and (b) R\"{o}ssler system. For the R\"{o}ssler system,
the equation is given by $\dot{x} =  -y -z, \dot{y}=x+ay,  
\dot{z}=xz-cz+b$ ($a=0.2, b=0.4$, and $c=8.5$) and the phase is
defined by $\tan^{-1} (y/x)$.}
\label{local}
\end{figure}

\begin{figure}
\caption{ Coupled Lorenz system. $\epsilon =1.0, N=300$.
Time evolution of $x_{i}$ is plotted in (a) (from $i=1$ to $i=300$) 
and in (c) (from $i=201$ to $i=300$).  In (b), local frequency gradient 
$\delta w=(|\omega_{i}-\omega_{i-1}|+|\omega_{i}-\omega_{i+1}|)/2$ is
calculated from the same data in (a). In (a) and (b), the brightness 
is proportional to the largeness of $\omega$ and $\delta w$.
In (d) and (e), $\omega_{i}$ is obtained from the regions
denoted by ``frozen'' and ``pulsating'' in (a):
(d) frozen (e) pulsating.  }
\label{coupled}
\end{figure}

\begin{figure}
\caption{ Probability distribution function of the local frequency gradient,
$\delta w=(|\omega_{i}-\omega_{i-1}|+|\omega_{i}-\omega_{i+1}|)/2$,
for frozen (solid line) and pulsating (dashed line) domains. }
\label{hist}
\end{figure}

\begin{figure}
\caption{ Forced Lorenz system. $\epsilon =1.0$  
(a) $\overline{\omega} - \Omega$ vs $\Omega$ 
(b) Inside the solid line, imperfect synchronization occurs for the
uncoupled single oscillator.  In Region I (shaded region),
the confinement is broken and dynamics like (d) or (f) occur. 
In Region II (unshaded region), confined two-phase domain structure 
such as (e) forms. (c) A closer view of (b): triangle - (d), square 
- (f), elsewhere - (e). In (d),(e) and (f), the brightness is 
proportional to $x_{i}$. } 
\label{forced}
\end{figure}

\begin{figure}
\caption{ (a) Chaotic orbit of a single Chua oscillator and 
(b) its instantaneous frequencies $\omega (t)$.}
\label{chua}
\end{figure}

\begin{figure}
\caption{ Coupled Chua oscillators ($N=300$).
Space-time diagram : (a) $\epsilon=1.0$ and (b) $\epsilon=5.0$,
the brightness is proportional to $x_{i}$. 
(c) Time series of $x_{i}$ at $i=20$, and $i=90$ in (b)}
\label{chua2}
\end{figure}

\begin{figure}
\caption{ Lattices of forced Duffing oscillators
($a=-1,b=1,c=-0.25$, $\omega=1$, and $N=300$).
Orbits of uncoupled Duffing oscillators for (a) $f=0.29$ and (c) $f=0.40$.
Binary representations of space-time evolution for the coupled lattice
are given for $f=0.29$ ((b) $\epsilon=1$)  and for $f=0.40$ 
((d) $\epsilon=5$, (e) $\epsilon=60$, and (f) $\epsilon=100$).}
\label{duffing}
\end{figure}

\end{document}